# Investigating the ranges of (meta)stable phase formation in $(In_xGa_{1-x})_2O_3$: Impact of the cation coordination

C. Wouters[1], C. Sutton[2], L. M. Ghiringhelli[2], T. Markurt[1], R. Schewski[1], A. Hassa[3], H. von Wenckstern[3], M. Grundmann[3], M. Scheffler[2], M. Albrecht[1]

[1]*Leibniz-Institut für Kristallzüchtung, Max-Born-Str. 2, 12489 Berlin*

[2]*Fritz Haber Institute of the Max Planck Society, Faradayweg 4, 14195 Berlin*

[3]*Felix Bloch Institute for Solid State Physics, University of Leipzig, Linnéstraße 5, 04103 Leipzig*

**We investigate the phase diagram of the heterostructural solid solution $(In_xGa_{1-x})_2O_3$ both computationally, by combining cluster expansion and density functional theory, and experimentally, by means of TEM measurements of pulsed laser deposited (PLD) heteroepitaxial thin films. The shapes of the Gibbs free energy curves for the monoclinic, hexagonal and cubic bixbyite alloy as a function of composition can be explained in terms of the preferred cation coordination environments of indium and gallium. We show by atomically resolved STEM that the strong preference of indium for six-fold coordination results in ordered monoclinic and hexagonal lattices. This ordering impacts the configurational entropy in the solid solution and thereby the $(In_xGa_{1-x})_2O_3$ phase diagram. The resulting phase diagram is characterized by very limited solubilities of gallium and indium in the monoclinic, hexagonal and cubic ground state phases respectively but exhibits wide metastable ranges at realistic growth temperatures. On the indium rich side of the phase diagram a wide miscibility gap is found, which results in phase separated layers. The experimentally observed indium solubilities in the PLD samples are in the range of $x=0.45$ and $x=0.55$ for monoclinic and hexagonal single-phase films, while for phase separated films we find $x=0.5$ for the monoclinic phase, $x=0.65-0.7$ for the hexagonal phase and $x \geq 0.9$ for the cubic phase. These values are consistent with the computed metastable ranges for each phase.**

**Introduction**

Solid-solutions of group-III sesquioxides ($Al_2O_3$, $Ga_2O_3$ and $In_2O_3$) show promise in designing new transparent *n*-type electrodes or active materials for optoelectronic applications because of the ability to tune the bandgap energies over large ranges (i.e., 3.6 to 7.5 eV) by varying the relative cation concentration[1–7]. Moreover, heterostructural solid solutions such as the group-III sesquioxides where the binary components differ in their ground-state structures, offer the possibility of controlling the crystal structures by tuning the composition. This provides an additional degree of freedom for materials design beyond the effects of chemical substitution. Indeed, several current technologies such as optoelectronics[8], water splitting[9] and piezo-electronics[8,10] use structural modification by alloying as a route to widen potential applications.

To achieve the synthesis of high-quality $(In_xGa_{1-x})_2O_3$ alloys, an understanding of the phase formation as a function of composition is required. This is not a straightforward task, as $Ga_2O_3$ adopts a monoclinic (β) structure with space group $C2/m$ with mixed four- and six-fold cation coordination[11], while $In_2O_3$ has a cubic bixbyite (c) structure with space group $Ia\bar{3}$ and only six-fold cation coordination[12]. Additionally, these compounds display a rich phase space, with several polymorphs existing for both $Ga_2O_3$[13,14] and $In_2O_3$[15–18] that are somewhat higher in energy than the ground-state structures. In $Ga_2O_3$, the α ($R\bar{3}c$)[19,20], γ ($Fd\bar{3}m$)[20,21], and orthorhombic κ ($Pna2_1$)[20,22] (in literature also sometimes referred to as ε phase) phases have all been reported in addition to the thermodynamically stable monoclinic β-phase. In addition to the polymorphs of the binary compounds, a hexagonal (h) $InGaO_3$ alloy phase (see Fig. 1) with space group $P6_3mmc$ has been observed experimentally[23]. The structure contains five-fold and six-fold coordinated sites for the cations in equal amounts and is the first polymorph to accommodate $Ga^{3+}$ in five-fold coordination. The theory work by Maccioni *et al.*[24] also showed the remarkable stability of this phase for $x=0.5$. Therefore, a key question becomes: what are the ranges of stability for the different phases upon alloying in the $(In_xGa_{1-x})_2O_3$ system?

There exist two recent theoretical works on $(In_xGa_{1-x})_2O_3$, based on DFT calculations with limited configurational sampling, whose results are not consistent with each other. According to Peelaers *et al.*[25], a lower bound temperature for the full miscibility of the monoclinic phase is $T=812K$, while Maccioni *et al.*[24] predict narrow stable ranges of $x<0.18$ for monoclinic, $0.4<x<0.6$ for hexagonal and $x>0.9$ for cubic compounds around $T=800K$. These works are not consistent with experimentally observed indium solubilities. Most experimental work is done on the epitaxial growth of β-$(In_xGa_{1-x})_2O_3$ thin films, and different indium solubility limits with different growth methods are found: $x<0.4$ by sol-gel [5], $x=0.04$ by metalorganic vapor phase epitaxy[26], $x=0.1-0.3$ by pulsed laser deposition depending on substrate and growth conditions[27,28] and $x=0.35$ by molecular beam epitaxy[4]. The work by Holder *et al.*[29] illustrated the possibility for huge metastable composition ranges in heterostructural alloys, which is something that also has to be considered for the $(In_xGa_{1-x})_2O_3$ system, since so many polymorphs exist.

To further explore this materials system, we will reevaluate the composition and temperature dependent phase diagram of $(In_xGa_{1-x})_2O_3$ heterostructural alloys through a joint computational and experimental investigation approach. In contrary to the existing theory studies, we employ a computationally efficient protocol to search the vast configurational space of substitutional alloys using first-principles based cluster expansion models (CE) to understand the *ab initio* thermodynamics of these crystalline mixtures. We compare the computed thermodynamic phase diagram with experimental results from thin $(In_xGa_{1-x})_2O_3$ films with lateral variation of the alloy composition grown by pulsed laser deposition that are analyzed by high-resolution transmission electron microscopy to resolve the phase formation on atomic scale. We find that the mixing enthalpy as a function of composition for the monoclinic and hexagonal phases is strongly governed by the preferred coordination environments of gallium and indium atoms and ordered structures are energetically favored. This leads to a different behavior of the configurational entropy for these phases, in contrast to the cubic bixbyite phase which can be treated as an ideal solid solution. The resulting phase diagram exhibits large metastable composition ranges that agree well with the experimentally obtained solubility limits.

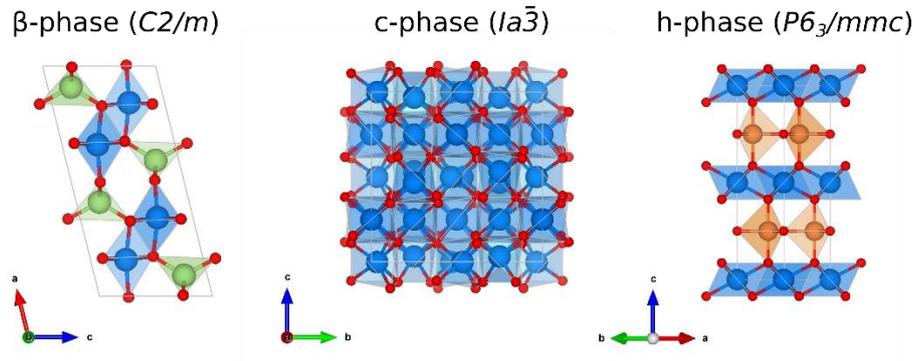

Figure 1. The lattice symmetries for the three sesquioxides considered in this work. The monoclinic, cubic bixbyite and hexagonal phases (referred to as β-, c- and h-phase in here) have only six-fold (blue), mixed four- (green)/six-fold, and mixed five- (orange)/six-fold coordinated cation positions with oxygen, respectively. Made with VESTA[30].

**Methodology**

**Experiment.** The samples investigated in this work are grown on (0001)-sapphire substrates using pulsed laser deposition (PLD) according to the continuous composition spread method[31]. In this approach, a two-fold segmented $(In_2O_3)/(Ga_2O_3)$ ceramic target is rotating synchronously with the substrate, which results in a thin $(In_xGa_{1-x})_2O_3$ film with a lateral continuously increasing average indium content. Energy dispersive X-ray spectroscopy (EDXS) and X-ray diffraction (XRD) 2ω-θ scans are performed along the compositional gradient of the samples to retrieve the indium content as a function of the position and the spatially resolved crystallographic properties at various indium concentrations. More details on these measurements and the PLD growth technique can be found in Ref.[28,31]. Transmission Electron Microscopy (TEM) samples are prepared in cross-sectional view along the $<1\bar{1}00>$ and $<11\bar{2}0>$ lattice directions of the sapphire substrate. Conventional and scanning TEM (STEM) measurements are performed with an aberration corrected FEI Titan 80-300 electron microscope operating at 300 kV, equipped with a high-angle annular dark-field detector (HAADF). Quantification of the indium concentration in the TEM samples is carried out by measuring the indium Lα and gallium Kα peak intensities using EDXS in STEM mode and analyzed using the Cliff-Lorimer method[32]. For the STEM-EDXS measurements, a JEOL JEM2200FS TEM with acceleration voltage of 200 kV is employed.

In the experimental results, we will distinguish between the 'globals' indium content $\tilde{x}$ as determined by SEM-EDXS in the scanning electron microscope and the 'local' indium content $x$ as determined by STEM-EDXS in the scanning transmission electron microscope. The SEM-EDXS is performed planarly and averages over μm-sized areas of the film. Therefore, possible local variations in indium content due to nanoscale phase separation are not registered in the measurement of the global indium content $\tilde{x}$.

Determination of the 'local' indium content $x$ is carried out in cross-section on the film by STEM-EDXS with nanoscale resolution, and will be important to determine the limits of indium/gallium incorporation in the separated phases.

**Computation.** For the computational part, the Gibbs free energy of mixing

$$\Delta G_l(x, T) = \Delta H_l - T\Delta S_l \quad (1)$$

of each ternary compound is computed for each lattice type $l$. In here, $\Delta H_l$ is the mixing enthalpy at 0 K computed according to the equation:

$$\Delta H_l = E[(In_xGa_{1-x})_2O_3]_l - xE[In_2O_3]_c - (1-x)E[Ga_2O_3]_\beta, \quad (2)$$

where $E[(In_xGa_{1-x})_2O_3]_l$ is the energy of the mixed system for phase $l$, and $E[In_2O_3]_c$ and $E[Ga_2O_3]_\beta$ are the energies of In$_2$O$_3$ in the c-phase and Ga$_2$O$_3$ in the β-phase, respectively. This definition gives an estimate of the energy relative to the stable binary phases. The entropy term consists of a configurational and vibrational contribution. The configurational entropy $\Delta S_{c,l}$ is calculated using the equation for the entropy of mixing of an ideal mixture

$$\Delta S_{c,l} = -N_l k_B (x \ln x + (1-x)\ln(1-x)) \quad (3)$$

where $N_l$ is the number of sites available for mixing in lattice type $l$ and $k_B$ is the Boltzmann constant. To calculate the vibrational entropy $\Delta S_v$ (independent of the lattice type) each atom (total $N$) is considered as a single-Debye-frequency oscillator and the mixture's Debye temperature $\Theta(x)$ is interpolated between those of the binary compounds, which are ~730K for Ga$_2$O$_3$[33,34] and ~700K for In$_2$O$_3$[35]. In that case, we can apply[36]

$$\Delta S_v = 3Nk_B((1+n)\ln(1+n) - n\ln(n)),$$

with $n(x, T) = \left[\exp\left(\frac{\Theta(x)}{T}\right) - 1\right]^{-1}$ as Planck's distribution. All quantities will be normalized to the number of cations.

As a first step, the lowest energy structures for all lattice types are identified using the cluster expansion (CE) method[37–39] which is a numerically efficient approach for examining the various configurational states of a specific lattice. The β-, c- and h-phase, of which unit cells are presented in Fig. 1, were found to be the lowest energy phases, and thus are further considered in this work. Then the mixing enthalpies at 0 K are recomputed using DFT according to Eq. 2, with the PBEsol exchange-correlation functional and the all-electron electronic structure code FHI-aims with tight settings[40]. We compared two GGA functionals (PBEsol and PBE) and selected PBEsol for this study because it gives the best accuracy for predicting lattice parameters in the group-III oxide systems. The average absolute difference between the volume (normalized by the number of cations) of the DFT-optimized structure and the ICSD[41] reported structures (cards #34243, #27431, #187791, #425685 and #187792) of five experimentally reported Ga$_2$O$_3$ and In$_2$O$_3$ polymorphs is lower by a factor of 3 for PBEsol compared to PBE. The PBEsol calculations are performed using a consistent 80-atom unit cell: 1x1x1 (i.e., the conventional unit cell), 2x2x1, and 2x2x2 supercell for the c-, β-, and h-phase, respectively. Once we have all $\Delta H_l$ values, the Gibbs free energies for various temperatures are easily calculated using Eqs. 1, 3 and 4. The convex hull analysis of the free energies as a function of composition used for determination of the temperature dependent phase diagram is performed using the qhull algorithm[42] in Python. The PBEsol calculations can be found on the NOMAD repository https://dx.doi.org/10.17172/NOMAD/2020.06.30-1 .

**Results**

A. Experimental

We have experimentally examined a set of $(In_xGa_{1-x})_2O_3$ TEM samples with different indium concentrations that were grown by the continuous composition spread method using PLD at growth temperatures of $T_g$ = 913-953K and a background oxygen pressure of $p(O_2) = 3 \cdot 10^{-4}$ mbar in the PLD chamber. These samples cover the compositional range $\tilde{x}$ = 0.0-0.9. TEM analysis of samples grown at $T_g$= 953K with $\tilde{x}$ = 0.1, 0.25 and 0.45 indicates that the β-phase is stable at these concentrations and the layers are single-phase. As an example, TEM images of the sample with $\tilde{x}$=0.45 (i.e., $(In_{0.45}Ga_{0.55})_2O_3$) are shown in Fig. 2(a), where the β-phase can be identified from both electron diffraction patterns and from the atomic pattern observed in STEM HAADF images (Fig. 2(a) bottom). The film is not single crystalline but nanometer-sized grains of the β-phase are formed, which are 60° rotated in-plane due to the hexagonal symmetry of the sapphire substrate. This explains the grainy contrast in the layer in the bright field image. The four different in-plane orientations of the monoclinic grains ([010], [0$\bar{1}$0], [132] and [1$\bar{3}$2]) are identified in the diffraction pattern. These results are consistent with the previous conclusions by von Wenckstern *et al.*[28] drawn from XRD data of the same film which indicated that the β-phase forms the major component over a composition of $0.0 \leq \tilde{x} \leq 0.5$.

For a higher indium concentration of $\tilde{x}$=0.55, a single-phase hexagonal $(In_xGa_{1-x})_2O_3$ layer grown at $T_g$=913K was observed in TEM, as illustrated in Fig. 2(b). The diffraction pattern shows the epitaxial in-plane relationship $[1\bar{1}00]_h \parallel [11\bar{2}0]_{sapph}$ between the hexagonal phase of the layer and the corundum sapphire substrate. Also here, the layer is not single-crystalline but the hexagonal phase consists of different grains with slight off-orientations with respect to each other. This is obvious from the diffuse diffraction spots and explains the short-range contrast variations in the bright field image. On the STEM image in the $[11\bar{2}0]$ orientation, some stacking faults can be observed indicating a defective hexagonal structure.

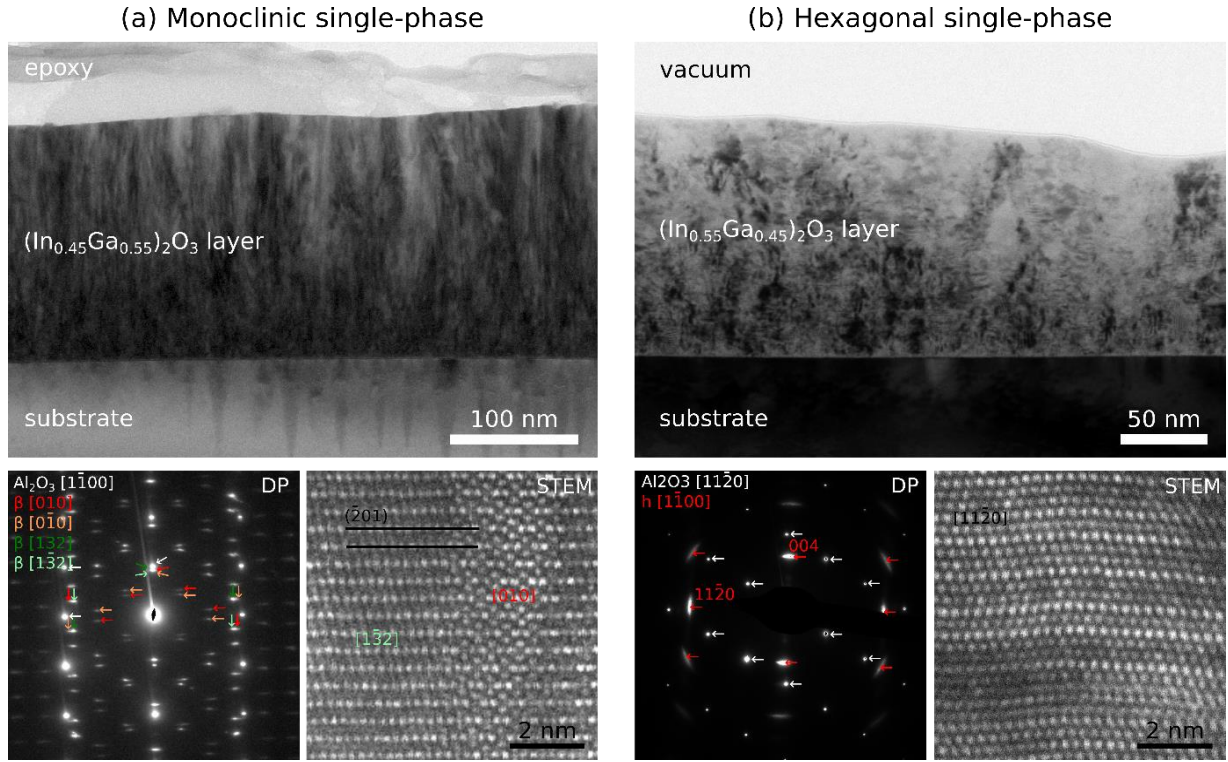

Figure 2: TEM images of single-phase (a) monoclinic $(In_{0.45}Ga_{0.55})_2O_3$ and (b) hexagonal $(In_{0.55}Ga_{0.45})_2O_3$ thin films on sapphire grown respectively at $T_g$=953K and $T_g$=913K. On top are TEM bright field images showing the morphology of the layer. The bottom left on each side shows the diffraction pattern of the $Al_2O_3$ substrate (white arrows) and the layer (colored arrows) together. For the monoclinic film, four different in-plane orientations can be recognized. Two of them are also identified in the STEM image, which shows the atomic pattern of an [010] and a [1$\bar{3}$2] grain and their common growth along the ($\bar{2}$01) planes. The STEM image of the hexagonal phase is taken in the [11$\bar{2}$0] orientation and presents some stacking faults.

For even higher indium contents, phase separation is consistently observed in the PLD layers. At $T_g$=953K, the relative X-ray peak intensities of the different phases in Ref.[28] indicate that the h-phase is the predominant phase for $0.5 < \tilde{x} < 0.7$, while the c-phase is dominating for $0.7 < \tilde{x} \leq 1.0$. TEM bright field and STEM HAADF analysis of this 953K sample at an indium content of $\tilde{x} = 0.75$ reveals a layered structure indicative of phase separation. The β-phase is observed at the interface to the substrate, followed by the h-phase, and the c-phase forming at the surface (Fig. 4). For this phase separated sample at $\tilde{x} = 0.75$, different local indium contents (measured using STEM-EDXS) of $x$=0.5 for the β-phase, $x$=0.65-0.7 for the h-phase, and $x\geq 0.9$ for the c-phase are found. Similar to the single-phase layers, the β- and h-phase present small grains, while the cubic phase is almost single-crystalline, as apparent by the more homogeneous intensity in the bright field image (Fig. 4(a)). As a result, for the β-phase and h-phase, the STEM-EDXS measurement is averaged over different nm-sized grains because of the unavoidable sample drift during the measurement due to beam-sample interaction. This is not an issue for the c-phase because of the bigger grain sizes.

The same types of STEM-EXDS measurements were done for a second phase separated layer, which was grown at $T_g$=913K and has a global indium content of $\tilde{x} = 0.8$. For this layer a grainy hexagonal phase was identified at the interface with a local indium content of $x\approx 0.7$. On top of that, a cubic $(In_xGa_{1-}$

$_x)_2O_3$ phase was found with a local indium content of $x \approx 0.9$. Contrary to the other phase separated sample, in this layer no monoclinic phase was found in the TEM measurements.

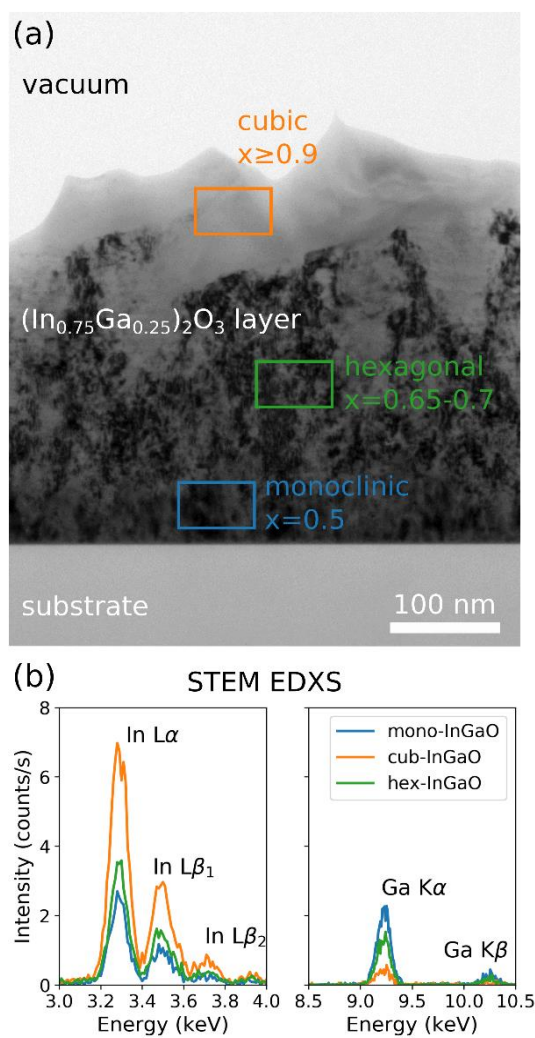

Figure 3: (a) TEM bright field image of a phase separated $(In_{0.75}Ga_{0.25})_2O_3$ thin film on sapphire substrate grown. The local indium content $x$ in the monoclinic, hexagonal and cubic regions is determined by the STEM-EDXS measurement shown in (b).

High-resolution STEM-HAADF (STEM Z-contrast) imaging in the TEM is used to determine the lattice site occupations of the gallium and indium atoms in the β-, h-, and c-$(In_xGa_{1-x})_2O_3$ alloys (Fig. 5). The respective indium contents for the images presented in Fig. 5 are $x = 0.45$, 0.55 and 0.9. Performing STEM-HAADF along a high symmetry zone-axis direction of the crystal provides information about the composition of atomic columns as their intensity scales roughly as $Z^2$ (Rutherford scattering), where Z is the average atomic number along the atomic column. This means that a higher intensity is observed when heavier indium atoms are present in a column. By imaging in the [010] ([110]) orientations of the β- (h-) phase, the well-ordered columns consist solely of four (five)-fold or six-fold coordinated cations. Individual four/five/six-fold coordinated cation columns in the STEM HAADF images have been identified by comparing the high-resolution image pattern to stick-and-ball models of the respective phases (see insets in Fig. 5). The black intensity line profiles shown at the bottom are averages of multiple line-scans extracted along two differently coordinated cation columns for the β- and h-phase. The higher STEM-HAADF intensity on the six-fold lattice positions (octahedral, blue ball in Fig. 5) compared to the four- and five-fold lattice sites for the β- and h-phase respectively, indicates in both cases a preference of indium for the six-fold coordination environment. To quantify this, we added the red intensity line profiles for simulated STEM-HAADF images of ordered monoclinic and hexagonal lattices with x=0.45 and x=0.55 respectively. In the monoclinic simulated structure, all indium atoms are randomly distributed on octahedral sites only; in the hexagonal simulated structure, all octahedral sites are occupied by indium and the remaining 5% of indium is randomly distributed on the tetrahedral sites. For the hexagonal phase we find a perfect agreement between experiment and simulation, while for the monoclinic phase the trend is similar but the intensity difference is slightly higher for the experimental structure. A possible explanation for this small discrepancy could be a locally higher indium content with all extra indium atoms on the octahedral sites as well.

In contrast, all cation lattice sites have the same six-fold environment in $c$-$(In_xGa_{1-x})_2O_3$, and therefore, only slight variations in the intensity are observed in the STEM-HAADF images due to the statistical incorporation of the gallium atoms. These results indicate that the occupation of the lattice sites by indium and gallium is consistent between the various phases across all examined compositions. These results provide experimental evidence for the strong energetic preference of $In^{3+}$ for an octahedral coordination environment.

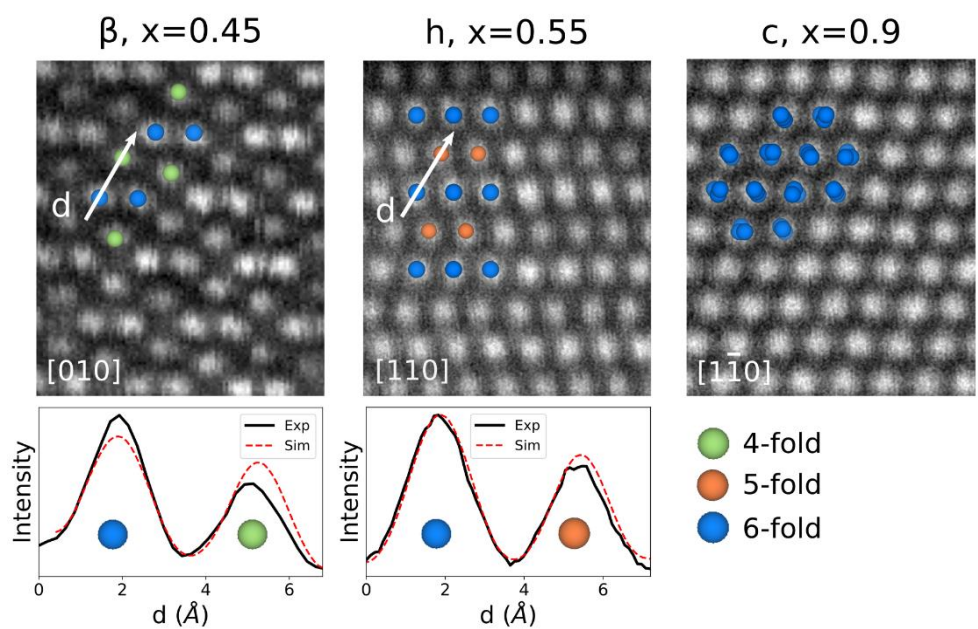

Figure 4. Experimental high-magnification STEM-HAADF images (several images summed to enhance contrast) of the β-, h-, and $c$-$(In_xGa_{1-x})_2O_3$ alloys with overlay of the stick-and-ball models of one unit cell without oxygen atoms. The lower plots show experimental HAADF intensity line profiles (black, multiple averaged) from the STEM images along the two differently coordinated atom columns in the β and h lattices, compared to the simulated profiles (red) for ordered structures. The intensity difference between the four-/five-fold and six-fold atomic columns, respectively, proves the preferential incorporation of the heavier indium to the six-fold lattice sites.

A. Computation

To understand the topology of the phase diagram, we start with a discussion of the PBEsol-computed ΔH values (per cation) of the lowest-energy structures at $T = 0$ K over the composition range $x = 0.0 – 1.0$ identified by the CE protocol (see Figure 5(a)). ΔH for the c-phase displays a concave parabolic shape over the whole composition range with a maximum at $x = 0.50$ because of the presence of only one six-fold coordinated cation site in this phase. Indeed, this energy surface is reminiscent of the classical energy of mixing with increasing concentration for an alloy with only one lattice site, e.g. $(In_xGa_{1-x})N$ in the wurtzite structure[29,43]. In contrast, the evolution of the ΔH for the β and h phases differs qualitatively from the classical behavior of the c-phase because these structures contain mixed four-/six-fold and mixed five-/six-fold coordinated cation sites with oxygen, respectively. For both phases, the lowest energy structures in the range $0.0 \leq x \leq 0.5$ correspond to those where indium is incorporated only into the six-fold coordinated lattice sites. For the β-phase, this results in a flat evolution in the mixing enthalpy with an increasing indium concentration, reaching a local maximum of ΔH = 0.035 eV/cation around $x = 0.34$ that decreases to ΔH = 0.024 eV/cation at $x = 0.5$. At this concentration, all gallium atoms occupy the four-fold coordinated positions, and all indium atoms occupy the six-fold positions, creating a long-range ordered structure where all indium and gallium atoms are in their preferred coordination environment. The h-phase is very unstable for $x = 0.0$ (ΔH = 0.167 eV/cation) and $x = 1.0$ (ΔH = 0.23 eV/cation) but displays a convex energy evolution and reaches a minimum at $x = 0.5$ (ΔH = -0.008 eV/cation). At this concentration, the lowest-energy structure corresponds to one with all gallium atoms on the five-fold sites and all indium atoms on the six-fold sites. Because of the equal amounts of six-fold/four-fold and six-fold/five-fold coordinated cation sites in the β- and h-phase, respectively, indium atoms for alloys with $x>0.5$ can only be incorporated into the four-fold/five-fold coordinated cation sites, which is energetically destabilizing and leads to the observed steep increase in energy for $x>0.5$. The ΔH data points are fitted to a single parabola for the cubic phase and two distinct parabolas for $x\leq0.5$ and $x>0.5$ for the β- and h-phase to reproduce the sharp edges at $x=0.5$. The fitted curves will be used for the further calculations of ΔG.

The large influence of the specific occupation site of gallium and indium on ΔH is highlighted in Fig. 6, where mixing enthalpies are calculated for about 100 random configurations of all three lattice structures at $x=0.5$. The mean effective coordination number (ECN)[44] of indium and gallium is determined for each configuration and plotted as a function of ΔH. Clear trends can be observed for the β- and h-phase of respective decreasing and increasing ECN for indium and gallium with increasing mixing enthalpy. The lowest energy configurations in both phases are the one with all indium atoms on six-fold sites and all gallium atoms on four- or five-fold sites. Going away from these ordered structures by displacing indium and gallium atoms to the other coordination environment drives a strong increase in the mixing enthalpy. This is in strong contrast with the situation for the c-phase, where the spread on the mixing enthalpies is very small since there is only one type of coordination site. This result explains the ordered β and h structures that we have observed in the STEM images in Fig. 4, where indium is sitting mostly on the octahedral lattice sites.

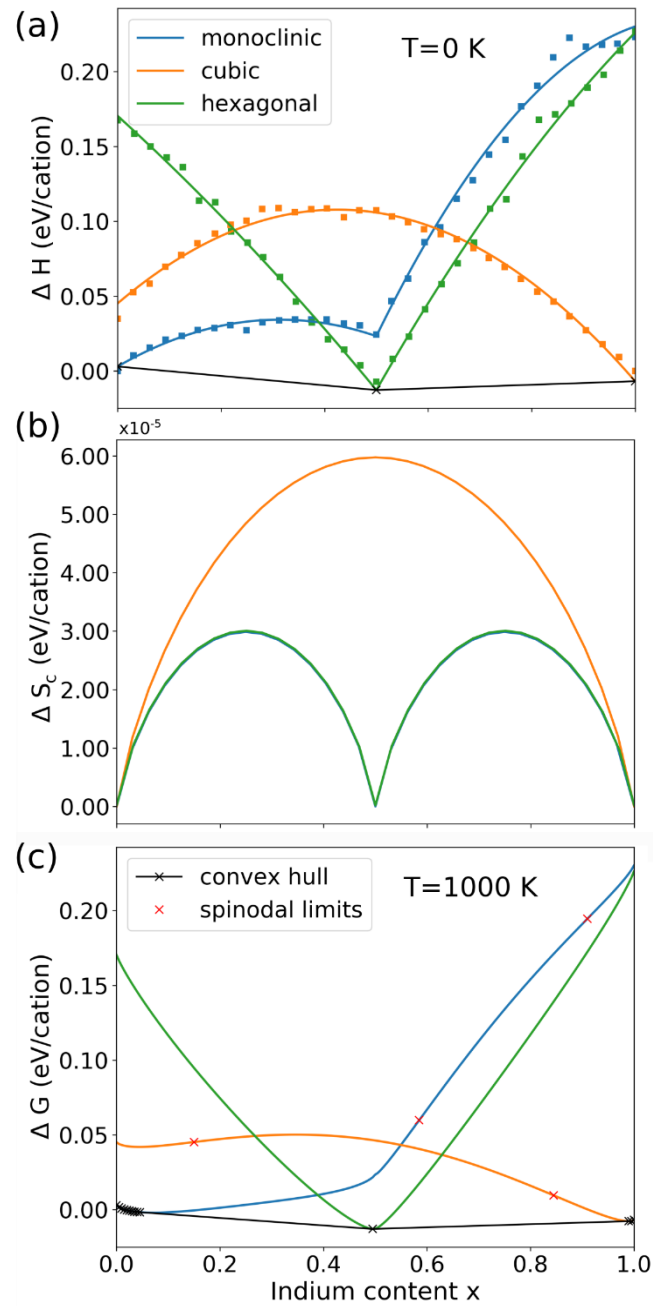

Figure 5: (a) Mixing enthalpies (ΔH) at 0K for the lowest energy configurations of the β-, h- and c-lattices calculated by DFT using the PBEsol functional with fit lines to the data. (b) Configurational entropy for the three phases, calculated as described in the text. (c) Free energies at T=1000 K as a function of indium content for the β-, h- and c-phase. The global convex hull at 0 K and 1000K is indicated by the black crosses and line. The spinodal limits at 1000K are indicated with red crosses.

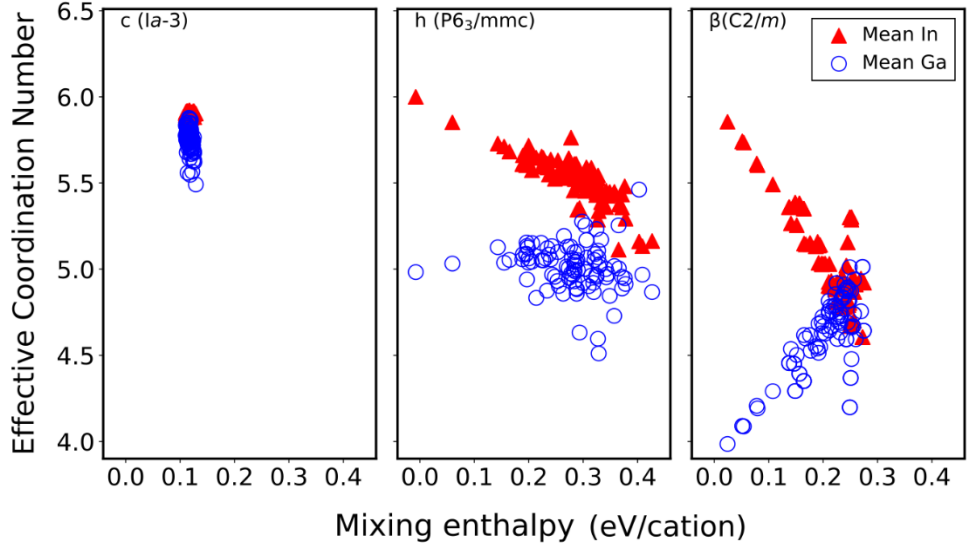

Figure 6: Mean effective coordination number of indium and gallium and corresponding ΔH values for about 100 randomly generated structures for c-InGaO$_3$ (left), h-InGaO$_3$ (middle), and β-InGaO$_3$ (right), i.e. $x=0.5$.

As mentioned in the Methodology section, the configurational entropy is calculated by Eq. 2 according to the ideal solution model which assumes a random distribution of the two components on the cation sites. When normalized to the number of cations, this results in the orange curve in Fig. 5(b). This we will apply for the cubic phase, since in this system indium and gallium are mixing on all sites with equal probability. Based on the results of Fig. 6 and the experimental results showing a strong preference of indium for the six-fold sites, mixing is not ideal in the monoclinic and hexagonal phase. Instead, we assume that for $x<0.5$ indium is only mixing on the six-fold sites, for $x=0.5$ β- and h-structures are highly ordered and for $x>0.5$ indium is mixing only on four-/five-fold sites in β- and h-(In$_x$Ga$_{1-x}$)$_2$O$_3$ respectively. This means that the entropy is zero at $x=0.5$, and that $N_l \rightarrow 1/2 N_l$ and $x \rightarrow 2x$ for $x<0.5$ and $x \rightarrow 2(1-x)$ for $x>0.5$ in Eq. 2, which gives the green/blue curve for the entropy in Fig. 5(b). The vibrational entropy, which ranges in between $10^{-7}$-$10^{-9}$ eV/cation depending on $T$ and $x$, is negligible compared to the configurational entropy and therefore not plotted here. The total entropy curves are applied to calculate the Gibbs free energies as a function of indium content for each phase for various temperatures. As an example, the free energy curves are plotted for $T=1000K$ in Fig. 5(c). A more detailed justification for the different treatments of the configurational entropy for the different phases is given in the Supplemental Material[45].

The thermodynamically stable phases and compositions can now be identified through the construction of the convex hull[46,47], which is comprised of a series of common tangent lines between the lowest free energy structures at various compositions on the free energy-composition surface. The convex hull at $T = 0K$ (i.e., without any entropic contributions) is given by the black line and crosses in Fig. 5(a), which indicates there are only three stable structures at $x = 0$ (β-phase), $x = 0.5$ (h-phase), and $x = 1.0$ (c-phase). The unstable mixtures in the range $0.0 < x < 0.5$ will phase separate into β-Ga$_2$O$_3$ and h-InGaO$_3$ ($x =$

0.5). For the indium-rich regime (0.5 < x < 1.0), the negative curvature indicates phase separation into h-InGaO$_3$ (x = 0.5) and c-In$_2$O$_3$ (x = 1.0). At higher temperatures, the free energy curves become more convex due to the -TΔS term, and more compositions will become stable, as can be seen for T=1000K in Fig. 5(c). The limiting compositions for stability for each phase and for temperatures up to 1400K are plotted in Fig. 7 by the dotted lines (binodals). The blue, green and orange filled regions then define the thermodynamic stable ranges for the β-, h- and c-phase respectively. The stable range is rather narrow, especially for the h-phase which is practically only stable at x=0.5 for all temperatures. The monoclinic stable window opens up more than the cubic one due to the rather flat ΔH curve for the β-phase for x≤0.5.

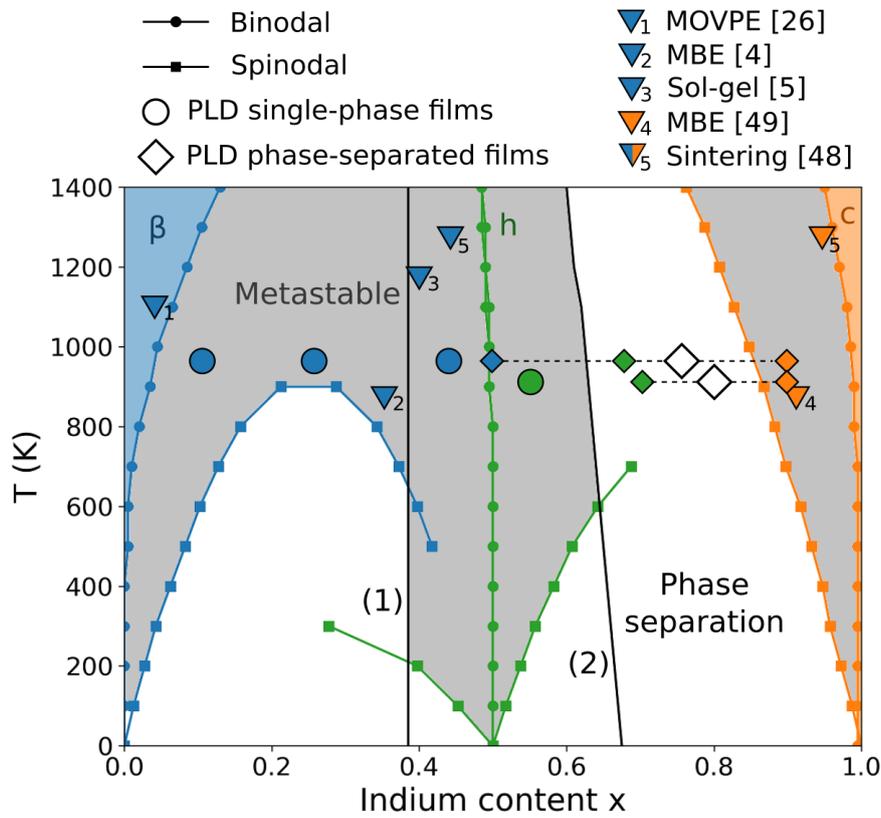

Figure 7: Computed temperature dependent phase diagram for (In$_x$Ga$_{1-x}$)$_2$O$_3$ including binodal and spinodal lines for the monoclinic (blue), hexagonal (green) and cubic (orange) phases. Thermodynamic stable composition ranges are color-filled, metastable ranges are grey, and for the white area below the spinodal lines phase separation is expected. The black vertical lines indicate the critical compositions where the lowest energy phase changes from (1) monoclinic to hexagonal and from (2) hexagonal to cubic. Experimental data points from PLD layers studied in this work (circles and diamonds) and from other growth/preparation methods found in literature (triangles) are added as symbols. The white diamonds denote the global compositions of the phase separated films.

Since epitaxial growth methods (like PLD e.g.) do not always operate at thermodynamic equilibrium, it is also interesting to define metastable ranges in the (In$_x$Ga$_{1-x}$)$_2$O$_3$ phase diagram. To do this, we follow the work of Holder et al.[29], who have constructed phase diagrams including metastable regions for heterostructural materials systems, like MgZnO. Compounds are labelled as metastable, meaning stable against small composition fluctuations, when the second derivative of the free energy curve is convex, i.e. $d^2G/dx^2 >0$. The zero-crossings of $d^2G/dx^2$ at 1000K are indicated by the red crosses in Fig. 5(c) for each phase (h-phase is convex for all $x$). These limiting compositions form the so-called spinodal in the phase diagram, plotted by the squared colored lines in Fig. 7 for each phase. Metastable compounds are then found in the region in the phase diagram between the binodal and the spinodal lines. The black lines in the phase diagram define the 'critical' compositions for a phase transition, i.e. where the lowest energy structure changes from (1) monoclinic to hexagonal and from (2) hexagonal to cubic. They additionally limit the metastable region for each phase separately. For growth temperatures above 900K a large metastable region on the gallium rich side extends up to indium contents around $x$=0.6, containing the monoclinic compound up to $x$=0.385 and the hexagonal compound for $0.385 \leq x \leq 0.6$. On the indium-rich side a miscibility gap remains up to temperatures higher than 1400K, where phase separation is expected into a hexagonal and a cubic alloy.

**Discussion**

As a comparison between the computational and experimental results, the experimentally determined compositions from single-phase (circles) and phase-separated (diamonds) PLD samples studied in this work have been overlaid on the computed phase diagram in Fig. 7. Additionally, literature data points (triangles) representing indium and gallium solubility limits respectively in monoclinic[4,5,26,48] and cubic[48,49] $(In_xGa_{1-x})_2O_3$ alloys grown or prepared by different methods have been added. A first observation is that almost all experimental compositions strongly exceed the narrow calculated thermodynamic stability ranges, and actually fit better to the calculated metastable windows. Only the allowed indium incorporation in MOVPE grown $\beta$-$(In_xGa_{1-x})_2O_3$ at $T$=1100K[26], which is the growth method closest to thermodynamic equilibrium, does agree well with the predicted thermodynamic limit. For sintered $(In_xGa_{1-x})_2O_3$ powders[48] that were heated for several days to reach equilibrium, the measured Ga solubility limit in $In_2O_3$ fits well to our calculation of the thermodynamic limit at 1275K, while the In solubility limit in $Ga_2O_3$ falls in the calculated metastable regime. We note that all the discussed samples were not grown/prepared under the exact same oxygen regime, which is another important factor influencing the solubility limit of the occurring phases besides the temperature but is not incorporated in our calculations.

The possibility to grow metastable monoclinic compounds over a large composition range up to $x$=0.5 follows from the rather flat behavior of its $\Delta H$ curve in this range (see Fig. 5(a)). This flat behavior in turn follows from the fact that indium can be accommodated in its preferred six-fold coordination environment for $0<x<0.5$. As a result of this, for increasing temperatures, the free energy curve will relatively quickly convert to a convex shape due to the entropy contribution. While for ideal mixtures, the spinodal is concave parabolic with a maximum at $x$=0.5 where there is maximum disorder, the monoclinic spinodal in this case presents this behavior over the range $0 \leq x \leq 0.5$ with the maximum at $x$=0.25 where there is maximal disorder on the octahedral sites.

At $x=0.5$, both the β- and h-phase display ordered structures as the lowest energy configuration and their mixing enthalpies reach (local) minima. Of the two, the hexagonal phase is slightly lower in energy and therefore the expected phase with a metastability range of $0.385 \leq x <\approx 0.63$ at realistic growth temperatures. Experimentally however, both phases are observed close to this composition range. The calculated energy difference between the β- and the h-phase in this critical region around $x=0.5$ equals 32 meV/cation (=12.8 meV/atom), which is on the order of the DFT accuracy found for similar oxide systems (24 meV/atom from Ref.[50]). Therefore, we could argue that both metastable phases are competing close to $x=0.5$ and both can be achieved.

The computed miscibility gap on the indium rich side stays up to temperatures above 1400K, because the ΔH curves of the h- and c-phase are much steeper there. The miscibility gap is well reproduced in experiment, as samples with global compositions of $\tilde{x}=0.75$ and $\tilde{x}=0.80$, and grown respectively at $T=953K$ and $T=913K$, indeed present phase separation. In this temperature range, the separated metastable phases according to calculation would be the h-phase with $x \approx 0.63$ and the c-phase with $x \approx 0.86$. This computed indium composition of the cubic phase agrees very well to the experimental compositions, both for our PLD layers and the reported MBE layer[49]. The computed indium limit of $x=0.63$ for the hexagonal phase is exceeded in our PLD layers, where we get a maximum incorporation of $x=0.7$. This could possibly be explained by the defective structure of the hexagonal phase in experimental films, which could allow for more indium to be incorporated.

**Conclusion**

In contrast to standard models of mixing, we have shown that interesting phases of heterostructural $(In_xGa_{1-x})_2O_3$ alloys could be achieved, because the different preferred coordination environments of the component elements are satisfied. The accommodation of indium in the preferred six-fold environment leads to low mixing enthalpies in the composition range $0 \leq x \leq 0.5$ for β-$(In_xGa_{1-x})_2O_3$ and a remarkable stability for h-$(In_xGa_{1-x})_2O_3$ close to $x=0.5$. This leads to large metastable windows in the gallium rich regime of the phase diagram for both phases. Indeed, it is experimentally confirmed that a large composition range of the monoclinic alloy can be grown ($x \leq 0.5$) as well as the hexagonal alloy at intermediate compositions up to $x=0.7$. On the indium rich side of the phase diagram, a miscibility gap remains up to high temperatures, which is confirmed in the experimental films which display phase separation. The c-phase, which has only one type of coordination site, is stable for $x \geq 0.9$ for growth temperatures around $T=1000K$, which fits well to the predicted metastable limit.

**Acknowledgements**

This work was performed in the framework of GraFOx, a Leibniz ScienceCampus funded by the Leibniz Association. This project has received funding from the European Unions Horizon 2020 research and innovation program (No. 676580: the NOMAD Laboratory and European Center of Excellence and No. 740233: TEC1p). This work was supported by European Social Fund within the Young Investigator Group "Oxide Heterostructures" (No. SAB 100310460) and partly by Deutsche Forschungsgemeinschaft in the Framework of Sonderforschungsbereich 762 "Functionality of Oxide Interfaces".